\documentclass[useAMS,usenatbib]{mn2e}
\usepackage{graphicx}
\usepackage{epsfig}
\usepackage{amsmath}

\title[Spectroscopic Study of HD192641]{Spectroscopic Study of the Long-Period Dust-Producing WC7pd+O9 Binary HD192641}

\author[L. Lef\`evre et al.]{L. Lef\`evre$^{1,2}$\thanks{E-mail: lefevre@astro.umontreal.ca,sergey@astro.wku.edu,
lepine@amnh.org, moffat,demers, skalkowski@astro.umontreal.ca, acker@astro.u-strasbg.fr, th@astro.ex.ac.uk, 
annuk@aai.ee, david.bohlender@nrc-cnrc.gc.ca, gyves@physics.mcgill.ca, hill@keck.hawaii.edu, nmorris2@uoft02.utoledo.edu,
 \mbox{dknauth@pha.jhu.edu}, sv@star.ucl.ac.uk }
, S. V. Marchenko$^{3}$, S. L\'epine$^{4}$, A. F. J. Moffat$^{1}$, A. Acker$^{2}$, 
\newauthor T. J. Harries$^{5}$, K. Annuk$^{6}$, D.A. Bohlender$^{7}$, H. Demers$^{1}$,  Y. Grosdidier$^{8}$,
\newauthor G. M. Hill$^{9}$, N. D. Morrison$^{10}$, D. C. Knauth$^{11}$, G. Skalkowski$^{1}$, S. Viti$^{12}$.\\
$^{1}$D\'epartement de Physique, Universit\'e de Montr\'eal and Observatoire du mont M\'egantic, C.P. 6128, Succ. ``Centre-Ville'',\\
Montr\'eal, QC H3C 3J7, Canada.\\
$^{2}${Observatoire de Strasbourg, Universit\'e Louis Pasteur, 11 Rue de l'Universit\'e, 67000 Strasbourg, France.}\\
$^{3}${Department of Physics and Astronomy, Western Kentucky University, Bowling Green, KY 42101-3576, USA.}\\
$^{4}${Department of Astrophysics, Division of Physical Sciences, American Museum of Natural History, 79th Street, NY 10024, USA.}\\
$^{5}${School of Physics, University of Exeter, Stocker Road, Exeter, EX4 4QL, United Kingdom.}\\
$^{6}${Tartu Observatory, 61602, T\~oravere, Estonia.}\\
$^{7}${Herzberg Institute for Astrophysics, National Research Council of Canada, Victoria, BC V9E 2E7, Canada.}\\
$^{8}${Ernest Rutherford Physics Building, McGill University, 3600 rue University, Montr\'eal, QC H3A 2T8, Canada.}\\
$^{9}${W.M. Keck Observatory, 65-1120 Mamalahoa Highway, Kamuela, HI 96743.}\\
$^{10}${Ritter Observatory, Dept. of Physics \& Astronomy, The University of Toledo, Toledo, OH 43606-3390, USA.}\\
$^{11}${Department of Physics and Astronomy, Northwestern University, Evanston, IL 60208.}\\
$^{12}${Department of Physics and Astronomy, University College London, Gower Street, London Wc1E 6BT.}\\
}

\begin{document}

\date{Accepted 2004 December 15. Received 2004 December 14; in original form 2004 October 11}

\pagerange{\pageref{firstpage}--\pageref{lastpage}} \pubyear{2004}

\maketitle

\label{firstpage}

\begin{abstract}
We present the results of an optical spectroscopic study of the
massive Wolf-Rayet binary HD192641=WR137. These 1986-2000 data cover the dust-formation
maximum in 1997. Combining all available measurements of radial velocities,
we derive, for the first time, a spectroscopic
orbit with period $4766 \pm 66$ days ($13.05$ $\pm$ $0.18$ years). The resulting masses, adopting i=67${^\circ}$, are $M_{O}$=$20\pm2$$M_{\odot}$ for the O component
and $M_{WR}$=$4.4\pm1.5$$M_{\odot}$ for the WR component. These appear, respectively, $\sim$ normal and on the low side for the given 
spectral types.
 Analysis of  the intense multi-site spectroscopic monitoring in 1999 shows that the {CIII} $\lambda5696$
and  {CIV} $\lambda5802/12$ lines have the highest intrinsic variability levels.
The periodogram analysis yields a small-amplitude modulation in the absorption troughs of the {CIV}
$\lambda5802/12$ and {HeI} $\lambda5876$ lines  with a period
of 0.83 days, which could be related either to pulsations or large-scale rotating structures as
seen in the WN4 star EZ Canis Majoris (WR6).
Wavelet analysis of the strong emission lines of {CIII} $\lambda5696$ and {CIV}
$\lambda5802/12$ enabled us to isolate and
follow for several hours  small
structures (emission subpeaks) associated
with density enhancements within the wind of the Wolf-Rayet star. Cross-correlating the variability
patterns seen in different lines, we find a  weak but
significant correlation between the varability in emission lines with different ionization potential, i.e. in lines
formed at different distances from the WR stellar core.
Adopting a $\beta$ wind-velocity law, from the motion of individual subpeaks we
find $\beta$ $\sim$ $5$, which is  significantly larger than the canonical value
$\beta$ $\simeq$ $1$ found in O-star winds.
\end{abstract}


\section{Introduction}

Population I Wolf-Rayet stars are highly evolved He-burning descendents of massive OB-stars, which
are assumed to have undergone rapid luminous blue variable (LBV) mass-loss episodes before reaching the WR stage 
\citep{lan1994}. They exhibit intense broad emission lines produced by highly-ionized
atoms \citep{cont2000}, which form  a hot, fast and dense
stellar wind \citep{ham1998}. Furthermore, recent observations \citep{mof1988,lep1999,lep2000} revealed that the winds are
highly structured, although the ``Standard Model" of a WR-star wind was based on the simplifying assumptions
of spherical symmetry, homogeneity, time independence and a monotonic velocity law for the wind \citep{hil1991,ham1991}.
The highly structured nature of the winds may be related to numerous small-scale  density enhancements embedded in
the outmoving WR wind. Once included in the framework of the ``Standard Model", the micro-structuring (clumping)
allowed one to: (i) produce much better fits of the emission line profiles
\citep[especially in the red-shifted  electron-scattering wings: ][]{ham1998}, (ii) bring theoretical 
spectral energy distributions closer to the observed infrared and radio fluxes \citep{nug1998}.

There are two broad classes of WR stars: WN, where  nitrogen lines
dominate, and WC (WO), with dominance of carbon (oxygen). These two classes may
then be further divided according to spectral ionisation related to wind temperature. Many of the cooler
WC subtypes are known to be prolific dust makers. There are two situations in which a WC star can produce dust:
(1) in a presumably single WC star of late spectral subtype (mainly WC9 and some WC8) and (2) in a WC+O binary
with a relatively long-period, often eccentric, orbit and no restrictions on the WC subtype . Currently
 seven WC stars are known to be episodic dust makers with periods or suspected periods
of several years. We refer here to WR19 (WC4), WR48a (WC8), WR70 and WR98a (WC9), and three WC7 stars
 WR125, WR137 and WR140 \citep{wil1995}. All of these are confirmed or suspected
 binaries with elliptical orbits and massive companions. They serve as unique laboratories, allowing one to study formation
 of dust in most hostile environments \citep{har2004,mar2002,mar2003,mon1999,mon2002,tut1999,wil2001}.

As one of the dust-producing prototypes,  WR137 \citep[$WC7pd+O9$,][]{huc2001} was frequently observed in the last two decades. The
initial discovery that WR137 forms dust during large IR outbursts \citep{wil1985} prompted various
investigators to obtain observations in 1997-1998 during the most recent outburst. One of these studies based on near-IR  HST/NICMOS2 images 
\citep[around the predicted periastron passage,][]{mar1999}, revealed dust clumps leaving the system. Different IR and UBV photometry
 campaigns yielded a $\sim$ 13 year modulation \citep{pan2000,wil2001} which was supposed to be related to the presence of a binary. 
WR 137 also shows a strong polarisation ``line effect'', possibly produced in a spatially anisotropic (flattened) wind \citep{har2000}.
Two of the best studied ``line effect'' WR systems, WR6 \citep{har1998,har1999} and WR134 \citep{har1998}, reveal the presence of periodically repeatable 
structures in their winds \citep[and references therein]{mor1997,mor1999}. We organised an intense multi-site spectroscopic monitoring
campaign of WR137 in 1999-2000 in order to improve the orbital parameters and study the structure of the WR wind 
and thus obtain a better understanding of the dust formation process.
Here we report on the results of this campaign and its integration into previous observations.

\begin{table*}
\centering
\begin{minipage}{140mm}
\caption{Summary of Observing Runs. \label{tbl1}}
\begin{tabular}{@{}lccccccc}
\hline
Site & telescope & UT dates & No. Spectra & Sp. Coverage & Sp. Dispersion & S/N \\
     &    (m)    &          &  & (\AA)        & (\AA /pix)     & (cont.) \\
\hline
 DAO  	& 1.8  & 08/1986-07/1999 & 41 	& 5300-6100    	& 0.75 	& 250 \\
 KPNO  	& 0.9  & 11-19/09/1999   & 58 	& 5600-6000 	& 0.10 	& 100 \\
 OHP  	& 1.5  & 01/1997-07/1999 & 50 	& 5200-6100	& 0.35 	&  60 \\
 OMM  	& 1.6  & 07/1991-09/1999 & 55  	& 5100-6600	& 0.65 	&  80 \\
 RO 	& 1.0  & 06/1999-09/1999 & 67    	& 5100-6100	& 1.50 	&  90 \\
 TO  	& 1.5  & 06/1999-03/2000 & 27   	& 5600-6000	& 0.80 	&  60 \\
{\bf Total}   &      &                 &  236      &		&	&     \\
\hline
\end{tabular}

\medskip
\medskip
\end{minipage}
\end{table*}

\section{Observations}

Spectroscopic CCD data were obtained from 1986 to 2000 at six different observatories
(see Table~\ref{tbl1}) with 1-2 m class telescopes: the Observatoire de Haute
 Provence (France), hereafter OHP; the Observatoire du Mont Megantic (Quebec), OMM;
 the Tartu Observatory (Estonia), TO; the Dominion Astrophysical Observatory (Victoria, Canada), DAO;
 the Ritter Observatory (Toledo, Ohio), RO; and the Kitt Peak National Observatory (Arizona), KPNO.


\section{Reduction and Analysis}

All spectra were extracted locally at each observatory using standard
routines (mainly de-biasing, flat-fielding and conversion to a one-dimensional spectrum). The remaining
calibration steps (wavelength calibration, continuum rectification, removal of the telluric spectrum) were performed
for all the spectra in succession using standard IRAF\footnote{IRAF is distributed by the National Optical Astronomy
Observatories, operated by the Association of Universities for Research
in Astronomy, Inc., under cooperative agreement with the National
Science Foundation.}
routines, thus  producing a  homogeneous data set. During the continuum rectification we made sure that all
the parameters, especially the wavelength windows assigned to a line-free continuum,  were identical for all the spectra,
although the much
higher resolution but limited spectral range of the KPNO spectra forced us to choose fewer continuum
windows (although similar ones for the overlapping range) for their rectification. Spectral samples used for the
determination of the continuum are shown in the top panel of Figure \ref{fig:f1}.

To be able to detect and measure subtle temporal variations in  the WR line profiles, we carefully removed the
telluric lines from all the spectra. Using normalized spectra of WR 137 taken at different air masses, we created a
template of the telluric lines, then divided the appropriately scaled and binned (to attain the same resolution
as the target spectrum)  template into each observed spectrum.
Additional iterative adjustment of the intensities and positions of the telluric lines allowed us to achieve an optimal removal.

\subsection{Assessing variations in the spectra}

\subsubsection{Temporal Variance Spectrum} \label{311}

Observing line profile variability in WR spectra, one may encounter
several types of intrinsic variations: a stochastic component related to the small-scale inhomogeneities in
the wind \citep{mof1988,lep1999}, as well as periodic variations on various timescales from hours
to weeks, presumably caused either by pulsation, rotation or an orbiting companion \citep{bra1996}.
To assess whether the variability patterns are intrinsic to the star or due to spurious features introduced
during the reduction process (mainly  associated with the continuum fitting and wavelength calibration),
we use the  ``Temporal Variance Spectrum'' (TVS) approach \citep{full1996,prin1996,mar1998}.
At each wavelength (index j) we calculate:
\begin{eqnarray}
TVS_{j} &=& \frac{1}{N-1} \sum_{i=1}^{N} (\frac{\bar{\sigma_{0}}}{\sigma_{i}})^2
 \frac{1}{\bar{S_{j}}}
  (S_{ij}- \bar{S_{j}})^2
   \; - \; \bar{\sigma}^2 \rho^2_j
\end{eqnarray}
where index i refers to a given spectrum (total number $N$), $(\frac{\bar{\sigma_{0}}}{\sigma_{i}})^2$ 
is taken to be $1$ ($\bar{\sigma_{0}}$ is 
the noise determined for the entire time series and $\sigma_{i}$ the noise determined
from continuum pixels in each spectrum), $S_{ij}$ is a spectrum
taken at time $t_{i}$ at the $j^{th}$ pixel or wavelength and $\bar{S_{j}}$ is the spectrum averaged over time (index i). 
The last term of the equation helps to eliminate the spurious details arising from  small positional errors introduced
during the
wavelength calibration: $\bar{\sigma}$ is proportional to the typical error of the radial velocities, and  $\rho_j$ is the
gradient of $\bar{S_{j}}$ with respect to the wavelength.

\begin{figure}
	\epsfxsize=7cm\epsfbox[100 170 470 650]{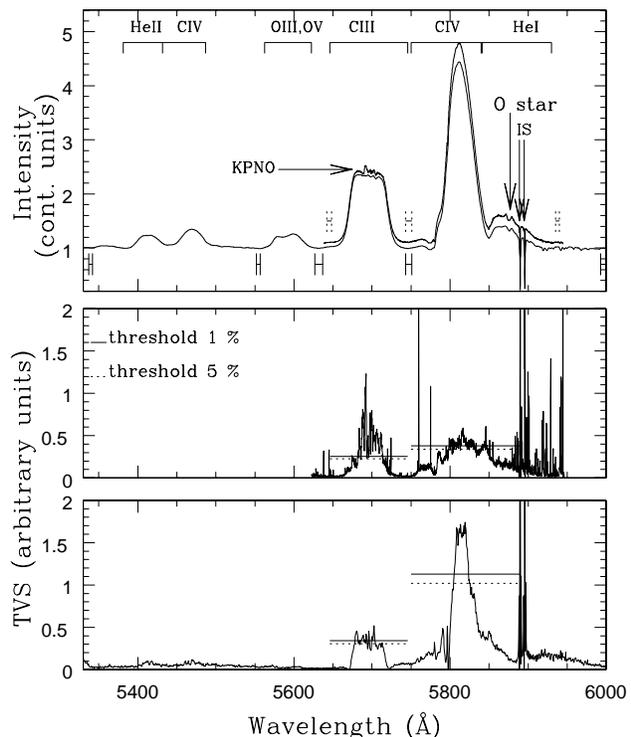}
	\caption{The upper panel shows the reference mean spectrum of WR137 (combined OHP,
	OMM, DAO, TO and RO data sets) with the superposed, vertically shifted (for clarity) mean spectrum from the
	KPNO set. The continuum windows used for both sets are displayed in the form of horizontal bars. The upper
	ones refer to KPNO data while the lower ones refer to OHP, OMM, DAO, TO and RO data.
	The middle panel shows the TVS for the KPNO data for September 11-19 1999, and the lower panel
	shows the TVS for the combined July 1999 sets from  DAO, OHP, OMM and TO.
        \label{fig:f1}}
\end{figure}

As defined,  the TVS  does not provide any detailed information about temporal characteristics
of the line profile variability. This can be remedied by calculating TVSs for isolated sub-sets.
 We have therefore calculated a TVS for each set of observations. We grouped the data sets in different
 ways to access different variability periods. Data sets were first grouped on a day-by-day basis,
 showing significant variability of the strong emission lines {CIII} $\lambda5696$  and {CIV} $\lambda5802/12$. Then the data were grouped in
 gradually larger intervals. The most representative sub-set for OHP, OMM, DAO, TO data (1999, July) is shown in
 Figure \ref{fig:f1} along with the TVS for the KPNO data (1999, September). The TVS for the RO data is similar in shape
 but the variations are less significant, as expected from the higher noise level of the data. It is not shown here for 
 reasons of clarity.

  Statistical behavior of the TVS follows a $\sigma_{0}^2 \chi_{N-1}^2$ distribution, hence thresholds
 (straight lines in our case) of a given statistical significance can be drawn for each TVS:
  $T=\sigma_{0}^2 \chi_{N-1}^2$, where N is the number of spectra and $\sigma_{0}$ refers to the noise level in the adjacent continuum.
In Figure \ref{fig:f1} one can see that the variations  in  the {CIII} $\lambda5696$ and {CIV} $\lambda5802/12$ lines
 are significant at the $99$\% level (i.e.,  exceeding the $1$\% threshold) for the combined OHP, OMM, DAO, TO data.
 
 It turns out that the observed profile variations (Fig. \ref{fig:f1} and other TVSs not shown here)
are practically always significant at the $99$\% level. 
The
spike-like features around the sharp interstellar Na D1/D2 lines at $\lambda$ 5890/96 are caused by combinations of spectra 
with different spectral resolutions (bottom panel) or difficulties in fitting the continuum (middle panel). Indeed, the RO TVS (not shown here) 
does not display these features as long as it is a combination of data from the same instrument and the continuum is well defined.

\subsubsection{Wavelet analysis} \label{331}

\begin{figure}
\epsfxsize=9.0cm \epsfbox[10 80 500 550]{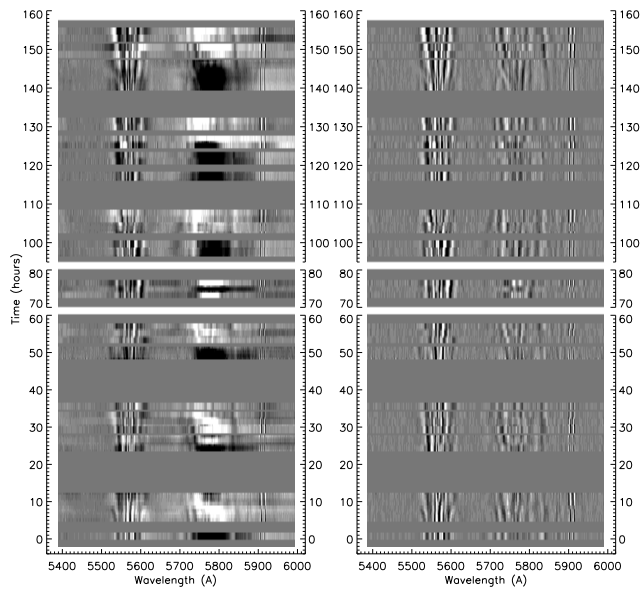}
\caption{Sequence of residuals of the strongest lines ({CIII} $\lambda5696$ and {CIV} $\lambda5802/12$)
 after the average spectrum of the 1999 multi-site run has been subtracted from each individual spectrum. The
residuals are displayed as a greyscale plot and stacked
sequentially from bottom to top. The white horizontal bands separate
spectra obtained on different nights. The grey horizontal bands represent missing spectra. This Figure represents nights
from July 21 to 31, 1999, where we have the highest density of data. The average spectrum is shown in Fig.\ref{fig:f3}. 
The left panel shows the residuals after subtraction, while the right panel
shows the same residuals after the large-scale variable components have been filtered out with
the power spectrum shown in Figure \ref{fig:f3}. 
Note the variability occurring in the form of narrow moving features in the {CIII} 
$\lambda5696$ emission line, and in the form of global line-strength 
variations in {CIV} $\lambda5802/12$ on the left panel. While the former is intrinsic to the star, the latter
is an artefact of the spectral rectification.
In the right panel, much of the (spurious) global
variability in the emission line strengths has been removed, leaving
only the signature of the intrinsic narrow moving
subpeaks. Variability patterns from the narrow subpeaks are quite
obvious in {CIII} $\lambda5696$ and {CIV} $\lambda5802/12$.\label{fig:f2}}
\end{figure}

\begin{figure}
\epsfxsize=8.0cm \epsfbox{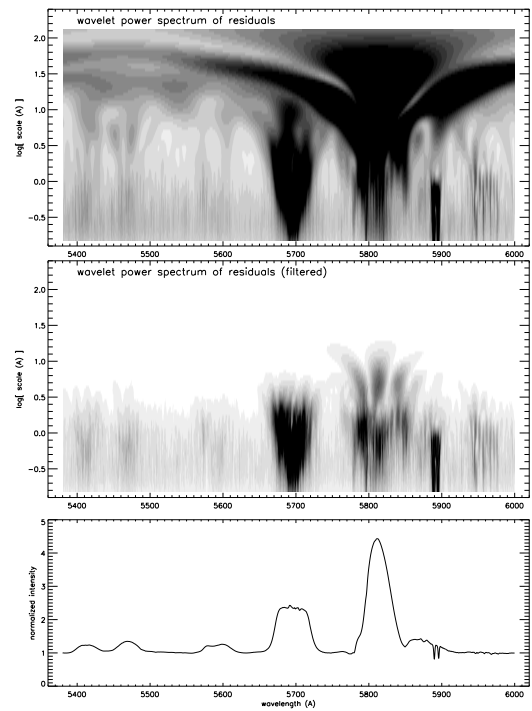}
\caption{Mean wavelet power spectrum of the residuals (top) and of the
wavelet filtered residuals (centre). The residuals yield a strong
response at small scales near the {CIII} $\lambda5696$ line, and at
larger scales near {CIV} $\lambda5802/12$, consistent with the
variability patterns observed in Figure \ref{fig:f2}. Filtering out the
large-scale (instrumental) components leaves only the signature of the
(intrinsic) narrow variable subpeaks. The wavelet power spectrum of
the filtered residuals yields a positive identification of narrow
emission features in both {CIII} $\lambda5696$ and {CIV} $\lambda5802/12$,
and suggests that narrow variable features are also present in {HeII}
$\lambda$5411, {CIV} $\lambda$5471 and {OV} $\lambda$5592 (grey patches).
\label{fig:f3}}
\end{figure}

In order to investigate the variability in the different spectral lines on a relatively small scale,
it is possible to use the wavelet techniques described in \citet{lep2000}. 
To perform this analysis of rapid spectral variations, a carefully computed mean spectrum must be subtracted from the individual
spectra.\\

The mean spectrum was constructed in steps. First, the sub-sets of the data (usually the rectified, telluric-free spectra
from the same observatory) were grouped into preliminary means. Then, these groups were cleaned from any deviating
spectra (comparing the preliminary mean to the individual spectra), forming the final mean spectra. The final mean spectra at similar
spectral resolution (OHP, OMM, DAO, TO) were used to form, via simple averaging, the template mean used in this section. Due to the
different spectral resolutions, the KPNO and RO data were first treated separately. We show that all the most prominent emission lines in
the {5400-6000\AA} range are variable, predominantly in the form of narrow emission subpeaks moving across the broad line profiles.\\

In Figure \ref{fig:f2} (left) we plot deviations from the average profile
(shown at the bottom of Figure \ref{fig:f3}) for spectra obtained during the
1999 multi-site campaign (OHP, OMM, DAO, TO, RO) for the {CIII} $\lambda5696$
and {CIV} $\lambda5802/12$ lines and only for the nights where the temporal
coverage is the best. Two types of
variability patterns are apparent in the residuals: (1) patterns
marked by narrow moving features as is most apparent in the {CIII}
$\lambda5696$ emission line, and (2) broad patterns that suggest that
the whole line is globally increasing or decreasing in strength, as is
most obvious in the {CIV} $\lambda\lambda5802/12$ emission
doublet.
The first patterns correspond to the
moving emission-line features seen in the spectra of many WR stars \citep{lep1999}.
The second pattern, on the other hand, is largely an artefact of the data reduction
procedure.
All spectra have been rectified to a pseudo-continuum level. The
pseudo-continuum is poorly defined in the regions with numerous blended, broad
emission features. The chosen pseudo-continuum sampling  produces relatively small, but,
nevertheless, easily detectable systematic
differences between the  groups of spectra (usually coming from different observatories).%

\begin{figure}
\epsfxsize=8.0cm \epsfbox{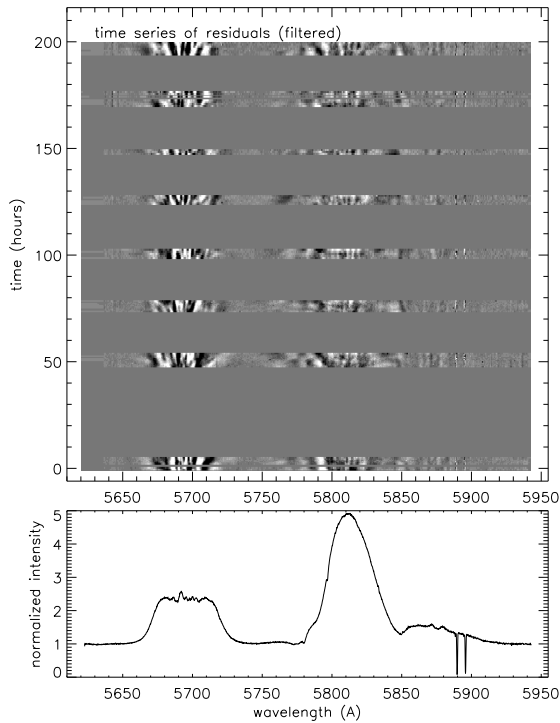}
\caption{The KPNO run: temporal sequence of residuals after the large-scale variable
components have been filtered out - compare to  Figure \ref{fig:f2}.
\label{fig:f4}}
\end{figure}

Fortunately, the variability of the intrinsic narrow moving features and that of
the global (spurious) line strength variations have very different
characteristic scales. It is thus possible to separate them
using wavelet analysis. In Figure \ref{fig:f3}, we display the
mean wavelet power spectrum {\em of the residuals}. The narrow
Discrete Wind Emission Elements \citep[DWEEs; see][]{lep1999} introduce
a significant response on scales of $\approx1$ \AA. Note the very dark
patch near {$5700$\AA} that is associated with the high-amplitude
DWEEs in {CIII} $\lambda$5696. Global line variations, on the
other hand, yield responses over much larger scales. 
The very large, dark patch near
{$5800$\AA} that extends to scales of several tens of {\AA} is the
signature of the large pseudo-continuum rectification artefact in
{CIV} $\lambda\lambda5802/12$. Note also the narrow dark patch
near {$5890$\AA} extending over a range of very small scales ($<1$
\AA): this corresponds to another artefact associated with the
interstellar absorption doublet. Because the spectra from different
observatories have slightly different spectral resolutions, subtraction of the
mean profile from the individual spectra introduces
residual features that are very apparent in Figure \ref{fig:f2}.

We use wavelet filtering to remove the rectification artefact from the
time series of residuals. Keeping only wavelet coefficients on the
smallest scales ($<4$\AA), we use the wavelet reconstruction theorem
to rebuild each individual residual spectrum. The mean wavelet power
spectrum of the filtered residuals is displayed in Figure
\ref{fig:f3}. A time series of the filtered residuals is
displayed in the right panel of Figure \ref{fig:f2}, and can be compared to the
original time series in the left panel of the same figure. One sees that all large scale variable features
have been filtered out. The narrow moving features in the {CIV} $\lambda\lambda5802/12$
emission doublet, which initially were completely drowned out in the
(spurious) global line-strength variations are now clearly visible. These intrinsic features
are weaker than those in {CIII} $\lambda5696$. This behavior mimics 
the variability patterns seen in emission lines of the single WC8 star
WR135 \citep{lep2000}.

A thorough examination of the filtered residuals and corresponding wavelet power spectrum also reveals what
appears to be a very weak signature from intrinsic narrow emission
features in the three prominent emission lines bluewards of {CIII}
$\lambda5696$. In Figure \ref{fig:f3}, darker patches can be seen
matching the locations of {HeII} $\lambda5411$, {CIV}
$\lambda5471$ and {OV} $\lambda5592$, indicating a significant
component of variable features at the scale expected for the narrow
emission-line subpeaks.

The same wavelet-filtering procedure is repeated for the spectroscopic
time-series obtained during the Kitt-Peak run. 
Again, random variations on large scales affect the {CIV}
$\lambda\lambda5802/12$ doublet, and these are an artefact of the
spectral rectification. 
Fortunately, these artefacts
occur on a scale ($>10$\AA) that is larger than that of the intrinsic
DWEEs, and they are easily filtered out. The resolution of the Kitt-Peak spectra ($0.1$ {\AA} per pixel) is
higher than that of the other sub-sets of data, hence one may obtain  a
clearer picture of the scale properties of the DWEEs. The behavior is
largely the same as noticed by \citet{lep1999}, producing, in the wavelet power diagram, the shape
of a symmetric triangle standing on its tip which marks the centre of the emission line, where the
DWEEs are narrower than near the edges of the line.
The time-series of the wavelet-filtered KPNO spectra is shown in Figure
\ref{fig:f4}. The general motion of the DWEEs, from the line
centre toward the line edges, is quite apparent in {CIII}
$\lambda5696$ and {CIV} $\lambda\lambda5802/12$, thus confirming the presence
of a strong variability as seen in the TVS analysis in section~\ref{311}. From the KPNO time-series, as for the previous one, the
DWEEs in {CIII} $\lambda5696$, if compared to  {CIV}
$\lambda\lambda5802/12$, do not give the impression that they
follow similar patterns. However, the doublet nature of the
{CIV} transition may confuse things considerably. Hence, we must explore
the possibility that the variability in the different
emission lines follows similar patterns. Such behavior would be
consistent with a model of the wind where the different emission lines
are formed at different depths in the wind on average, depending on
their excitation potential, but with a significant
overlap between the emission regions \citep{her2000}. Therefore, line profile variations caused by
outmoving clumps would be expected to show up in all the lines at similar
times \citep{lep1996,lep2000}. This will be discussed in detail in section~\ref{322}.

\subsubsection{Periodograms}

\begin{figure}
	\epsfxsize=8cm\epsfbox[50 190 550 620]{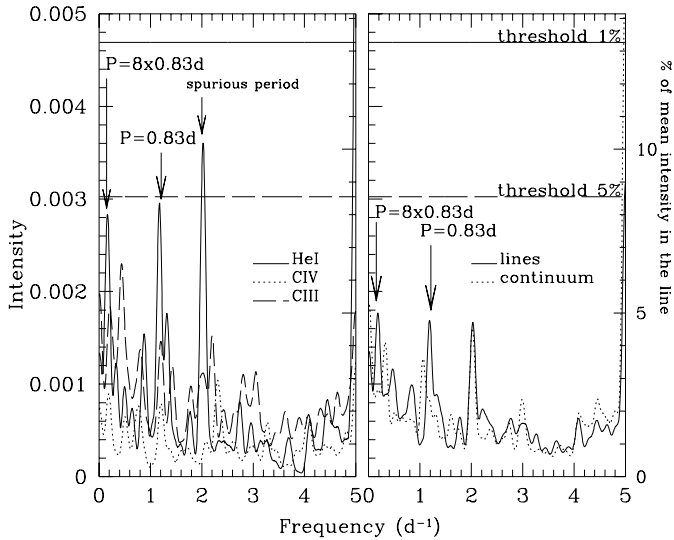}
        \caption{Kitt Peak Observatory periodograms reduced to one dimension. 
	The average periodograms for the P Cygni absorption part of the three main lines ({CIII} at 5670-5692, {CIV} at 5755-5776 
	and {HeI} at 5838-5848) are shown in the left panel while the average periodograms of the two main lines ({CIII} at 5670-5720 and 
	{CIV} at 5785-5835) along with the continuum ($\lambda \lambda$ 5645-5665,5735-5755,5900-5930) are shown in the right panel. The false alarm probability
	 threshold levels are also indicated.
        \label{fig:f5}}
\end{figure}

Inspired by the positive results of the TVS analysis, we also searched
the spectra for periods using the periodogram approach of \citet{sca1982}
combined with the cleaning algorithm of \citet{rob1987}.
Following \citet{sca1982}, we use $N_{0}=N/2$ frequencies $\omega$, noting that the number of
observations, N, could be different  for the TVS and periodogram samples. This defines the
corresponding false alarm probability thresholds: $T\approx {ln(N_{0}/p_{0})}$,
where $p_{0}$ is the false-alarm probability, usually set at 0.01-0.05. The cleaning procedure
effectively suppresses the false periodicities arising from the non-regular character of the observations,
the most prominent ones coming from the daily sampling bias of the data sets (see Fig. \ref{fig:f5}).
Although we found no periodic behavior in the spectra from OHP, DAO, TO, RO or OMM, this method yielded a small-amplitude, 
0.83 $\pm$ 0.04-day period in the KPNO spectra, searching in the range P=0.2-30d. Probably, both the high homogeneity of this
particular set, as well as its high spectral resolution helped to detect the low-amplitude signal.\\

We use wavelet filtering to isolate two different components of the variability: {\it small-scale} ($\le$ $4$ \AA) and {\it large-scale} 
($>$ $4$ \AA) structures (methods described in section~\ref{331}) in order to distinguish between real variability and artefacts of the spectral rectification.
The presence of this 0.83d period in the {\it large-scale structures} (i.e. associated with overall variations of the continuum caused
by errors in its fit) would invalidate its existence while its appearance in the {\it small-scale structures} would suggest that it is
real.\\

We found no significant signals in the {\it wavelet filtered} (either small- or large-scale) periodograms calculated from averaged
fluxes of the emission parts of the three major lines: {CIII}  (5670-5720 domain), {CIV} (5785-5835) and {HeI} (5865-5885). 
On the other hand, the periodogram of the 
 P Cygni absorption parts shows a clear peak at 0.83d for {CIII} while the peaks are diminished for the {CIV} and {HeI} lines. Moreover,
 the 0.83d signal is absent from the {\it large scale} structures in the {CIV}, {CIII} and {HeI} lines of the same periodogram.   
Combining the results from Fig.\ref{fig:f5} and the wavelet filtered periodograms, we conclude that: (1) there is no periodic signal in the emission
parts as suggested by the wavelet filtered periodograms of the lines; 
(2) there is a periodic signal at 0.83d, but it can be seen {\bf only} in the small-scale structures in {CIII} $\lambda5696$ 
(it is amplified by the filtering) and in the absorption part of {HeI} $\lambda5876$ (Fig. \ref{fig:f5}, left panel). Note that,
though the unfiltered signal in {CIII} $\lambda5696$ falls much below statistically significant levels, wavelet filtering allows one to increase the period's
visibility rather dramatically. This different behaviour of the {CIII} line is to be expected as it has no discernible P Cygni trough, and all the
variability seems to be concentrated there. It is no surprise either to find no 0.83d signal in the {CIII} line in the unfiltered data: this signal
 shows only in the filtered {\it small-scale} set, being restricted to the blue wing of the {CIII} line profile, roughly located at the same place as
  the {CIV} and {HeII} counterparts.

\begin{figure}
	\includegraphics[width=84mm]{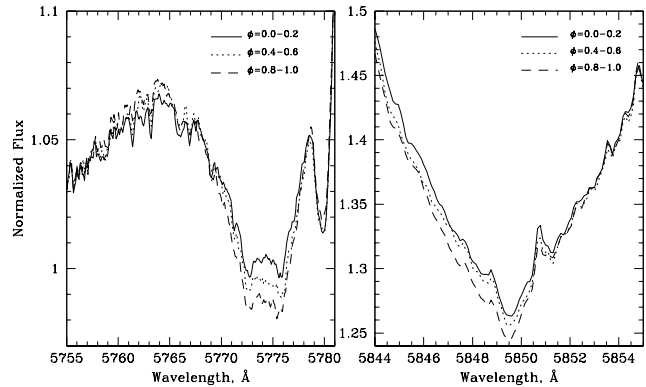}
        \caption{Absorption troughs of {CIV} $\lambda\lambda5802/12$ (left)
	and {HeI} $\lambda5876$ (right) phased (P=0.83d, $T_{0}$=HJD 2480198)  with the 0.83d period found exclusively in the KPNO data.
	Different lines depict average profiles calculated for the given phase bins.
        \label{fig:f6}}
\end{figure}

\begin{figure}
	\epsfxsize=8cm\epsfbox[10 300 520 490]{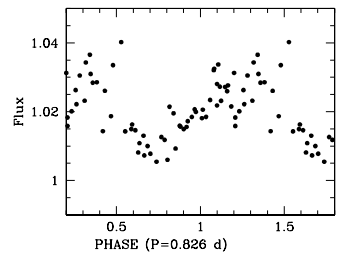}
        \caption{Averaged flux in the 5750-5780 {\AA} range of the {CIV} $\lambda\lambda5802/12$ line versus
	phase (P=0.83d; $T_{0}$=2450198d)  for the KPNO data.
        \label{fig:f7}}
\end{figure}

We show the phase-binned profiles of {CIV} and {HeI} in Fig.\ref{fig:f6} and an average flux coming from the P Cygni
absorption trough of {CIV} in Fig. \ref{fig:f7}. Both the amplitude ($0.5$\%-$1.5$\% of the continuum)
 and period of this variation are small compared to similar cases as seen for two other single WR stars with similar large-scale periodic
 spectral variations (WR134 and WR6), where the amplitude is $5$\%-$10$\% of the continuum and the periods are 2.31d and 3.76d,
 respectively \citep{mor1997,mor1999}.
 Note that, by themselves, the significance levels plotted in Figure \ref{fig:f5} are uncomfortably low.
 This may be related to the relatively high level of the noise introduced by the adopted procedure of
 rectification: note that, while calculating the respective values, we used the unfiltered KPNO spectra.
 However, Figures \ref{fig:f6} and \ref{fig:f7} lend strong support to the 0.83-day periodicity. Note that these periodic 
 variations show coherent patterns in two regions separated by $\sim 50${\AA} ($\sim 250$ resolution elements). 
 We further discuss the 0.83$\pm$ 0.04d period in section~\ref{disc}.

\subsection{Spectral-line variability} \label{bozomath}

\subsubsection{The $\beta$ wind-velocity law}

\begin{figure}
	\epsfxsize=8cm\epsfbox[40 150 520 690]{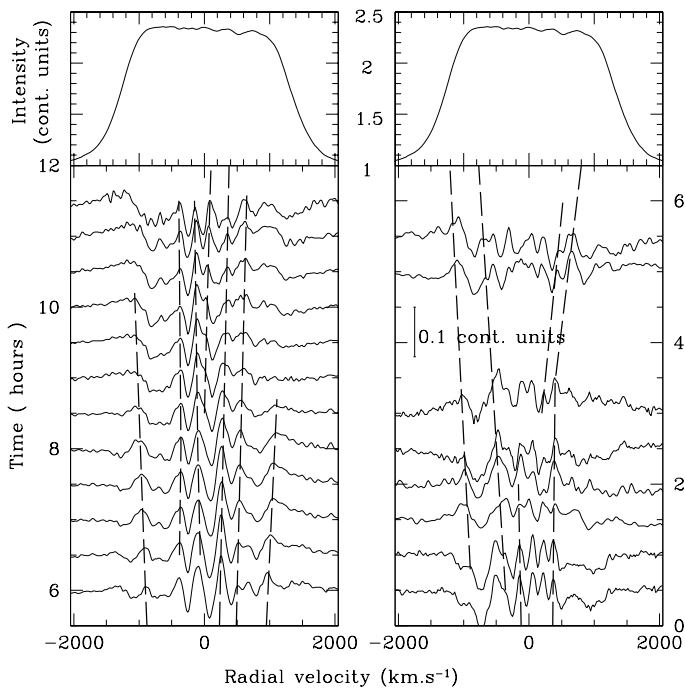}
	\caption{Clump evolution during two nights: left panel - 21/07/1999; right panel - 22/07/1999.
	The mean profile for all runs for the {CIII} $\lambda$5696 line is shown in the upper panels. The bottom panels show montages of the 
	deviations from the mean profile. In this plot we trace only the longest-living structures (straight dashed lines). However, 
	practically all moving features were chosen for further analysis (Fig. \ref{fig:f9}).
        \label{fig:f8}}
\end{figure}

\begin{figure*}
  \begin{center}
    \leavevmode
       \epsfxsize=12cm\epsfbox[70 190 500 620]{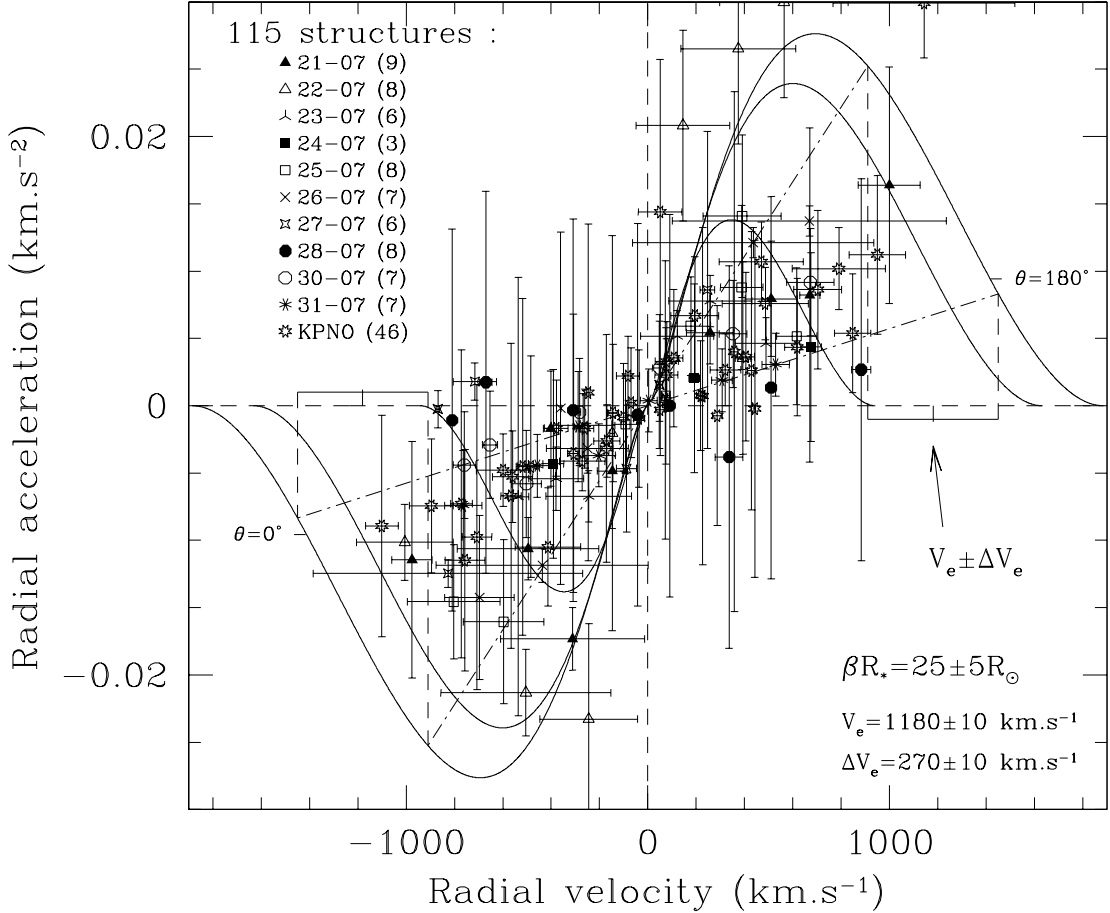}
       \caption{Radial acceleration versus radial velocity for detected blob trajectories from the DAO, OMM, OHP, RO, TO spectra of {CIII} 5696 for the period 
	of July, 1999. 
	The horizontal bars span the range of observed RVs while the vertical bars represent ${\pm}$ the standard deviations for the fitted accelerations. 
	A few points, due to measurement errors, fall into ``forbidden'' regions, i.e. on the back-side of the wind coming toward us and vice versa, providing wrong-sign
	accelerations. Curves based on eq. (4) represent angles to the line of sight from $\theta$=0 to 180 degrees in steps of 30 degrees, left to right, 
	bottom to top. The dash-dotted lines represent the inner and outer limits of the {\it ler} in velocity ($V_{e} = 1180 \pm 10$ $km\,s^{-1}$ and 
	$\Delta V_{e} = 270 \pm 10$ $km\,s^{-1}$ for the {CIII} $\lambda$5696 line) calculated according to \citet{lep1999} for $\beta R_{\star}$ 
	$\sim$ $25 R_{\odot}$. The two zones delimiting the {\it ler} in velocity space are indicated by horizontal bars.\label{fig:f9}}
  \end{center}
\end{figure*}

\begin{table*}
\begin{minipage}{175mm}
\caption{Stellar parameters of HD192641. \label{tbl2}}
\begin{tabular}{cccccccccc}
\hline
distance & $R_{\star}$ & $\beta$  & $v_{\infty}$   &  & $M_{WR}$ & $M_{O}$  & $M_{WR}$/$M_{O}$ & P & $L_{X}$ \\
(kpc) & ($R_{\odot}$) &  & ($km\,s^{-1}$) &  & ($M_{\odot}$) & ($M_{\odot}$) &  & years &  (erg $s^{-1}$) \\
\hline
 1.82$^a$ & 4.5 $\pm$ 2.5$^a$ & 5.6 $\pm$ 1.2$^b$ & 1885 $^a$ & & 4.4 $\pm$ 1.5$^b$ & 20.0 $\pm$ 2.0$^c$ & 0.22 $\pm$ 0.07 $^b$ & 13.05 $\pm$ 0.18 $^b$ & 0.25 $\pm$ 0.14 $10^{32}$$^d$ \\
\hline
\end{tabular}

\medskip
(a) \citet{nug2000},\citet{koe1995}, $v_{\infty}$ close to 1900 $km\,s^{-1}$ from \citet{huc2001}; (b) this paper; (c) \citet{vac1996},\citet{gie2003}; (d) \citet{pol1995}.\\
\end{minipage}
\end{table*}

We then examined the rapid spectral variations in order to try and constrain the $\beta$ wind-velocity law. To achieve this goal, the mean spectrum mentioned
in section~\ref{331}, was subtracted from the individual (unfiltered) spectra. 
Figure \ref{fig:f8} displays examples of the spectral
variability of the {CIII} $\lambda5696$ emission line  for the nights of July 21 and July 22,
1999. With the removal of the mean profile (top panels),
it is much easier to follow relatively weak emission details travelling on top
of the broadest lines \citep[for similar results in other stars see][]{gro2001,lep1999}.
As can be seen in Figure \ref{fig:f8}, individual subpeaks can be followed typically for several hours.
It is customary to relate the small emission peaks to small-scale density enhancements (``clumps" or ``blobs")
moving with the WR wind. Blobs are assumed to follow a $\beta$ wind-velocity law like the rest of the wind
(in fact they may constitute the wind as a whole):
\begin{eqnarray}
v(r) &=&   v_{\infty} (1-\frac{ R_{\star}}{r} )^{\beta}.
\end{eqnarray}
Their acceleration  is
\begin{eqnarray}
a(v) \equiv v \frac{dv(r)}{dr} &=& \beta \frac{v^{2}}{R_{\star}} [(\frac{v}{v_{\infty}})^{\frac{-\beta}{2}}- (\frac{v}{v_{\infty}})^{\frac{\beta}{2}}]^{2}.
\end{eqnarray}
The latter is equivalent for $\beta$ $^{>}_{\sim}$ 2 to \citep{lep1999}:
\begin{eqnarray}
a(v)={[v ln(\frac{v}{v_{\infty}})]^2}/{\beta R_{\star}}.
\end{eqnarray}
Assuming  $v_{\infty}=1885$ $km\,s^{-1}$ and $R_{\star}=4.5$ $R_{\odot}$ for WR137 (see Table~\ref{tbl2})
one can obtain an estimate of  $\beta$.

Following individual subpeaks (Fig. \ref{fig:f8}), one can draw straight lines which have a slope proportional to
the radial acceleration (i.e. the actual acceleration projected by an angle $\theta$ onto the line of sight) of the moving features.
Thus, measured accelerations (${a_{R}}$) along with the derived radial velocities ($v_{R}$) are plotted in an acceleration-velocity diagram 
(Fig. \ref{fig:f9}), which is traditionally used to constrain parameters of the $\beta$ wind-velocity law.
 The RVs of the subpeaks have been measured by fitting gaussian profiles to the moving peaks, assuming $\sigma(RV)\sim 50$ $km\,s^{-1}$
 for the blobs in each spectrum. The least-square fitting of a straight line to the positions of the peaks
provides estimates of the standard deviations for the acceleration.
Due to the exceptional temporal coverage of the two nights represented in Figure \ref{fig:f8}, it was possible to follow subpeaks even in the
regions corresponding to the steep flanks of the {CIII} $\lambda5696$ line.
These features move with  greater radial acceleration than the blobs originating near the centre of the line, due to projection effects, i.e. one observes
{${a_{R}}=a(v)cos\theta$} and $v_{R}=v(r)cos\theta$ at $\theta$=const. for each subpeak, such that those at line centre have $\theta$ $\simeq$ 90$^{\circ}$,
 and those near line edge have $\theta$ $\rightarrow$ 0$^{\circ}$ on the blue side or $\theta$ $\rightarrow$ 180$^{\circ}$ on the red side.

Figure \ref{fig:f9} shows that most of the components satisfy the expected $a_{R} v_{R} \ge 0$ relation,
thus confirming the initial assumption that the moving peaks are related to radially accelerated structures leaving the star.
Note that measurements with excessively large errors in acceleration, $\sigma_{a_{R}} \ge$ $0.02$ $km\,s^{-2}$, have not
been drawn in Figure \ref{fig:f9} to avoid clutter.
The 142 spectra (10 nights of observation at OHP, OMM, DAO, TO and RO during the period of 1999 July 21-31, except for July 29 
when only one spectrum was available, plus 8 nights of obervation at KPNO during the period of September 1999) result in
$\beta R_{\star} = 25$ $\pm$ $5$ $R_{\odot}$, i.e. $\beta$ $\simeq$ $5$, which is reasonably close to the range of values determined for WR137 by 
\citet{lep1999} ($\beta R_{\star}$ $\sim$ 35-90 $R_{\odot}$). 

\begin{figure}
	\includegraphics[width=84mm]{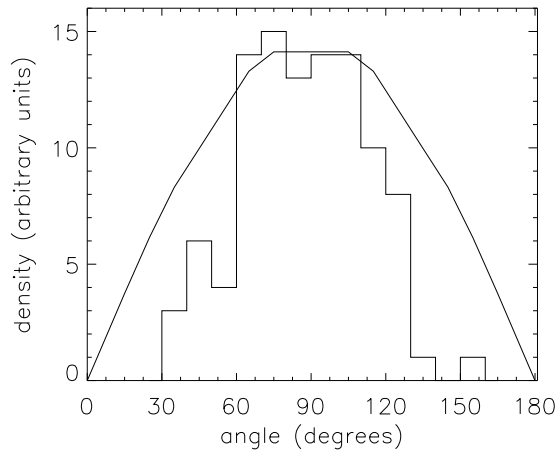}
        \caption{Actual density of {\it blobs} (histogram) compared to theoretical density (continuous line) versus angle of 
	projection. See text for details. The relative deficit of observed blobs vs. theoretical number of fast structures towards the wings is 
	likely due to the difficulty of measuring blobs in the wings of the {CIII} $\lambda$5696 line.
        \label{fig:f10}}
\end{figure}

Note that the lower-acceleration structures outnumber the higher-acceleration ones. One may find an explanation for this in Figure \ref{fig:f10},
which shows the actual density of blobs compared to their theoretical density (N($\Omega$) $\sim$ $sin(\theta)$, assuming constant spherically-symmetric
density) versus the angle with respect 
to the line of sight ($\theta$). 
The theoretical density maximum has been adjusted by a multiplicative factor to the maximum of the observed distribution. As can be seen, 
both observed and theoretical densities fall off as one goes away from $\theta$=90$^{\circ}$. However, there seems to be 
a real deficit in the observed density of blobs coming toward us ($\theta$ $\sim$ 0$^{\circ}$) or going away from us 
($\theta$ $\sim$ 180$^{\circ}$) compared to the expected density. 
This can be explained by the fact that events responsible for the high acceleration structures normally occur far from the centre of 
the spectral line, thus being difficult to observe and track: (1) the corresponding features move rapidly and require good temporal
coverage and (2) they are hard to recognize and trace because of their positions  and broader/flatter profiles
superposed on the steep slopes of the line.
In Figure \ref{fig:f9}, the limits of the {CIII} $\lambda5696$ line emission region ({\it ler}) have been calculated according to \citet{lep1999}.
Note that some points are located well outside the central cluster of points in Figure \ref{fig:f9}, i.e. outside the {\it ler} in the ``allowed" regions. 
These ``outliers" might result from the superior temporal coverage provided by the more complete data from July 21 and 22 and some nights
from the KPNO dataset. This may have
allowed one to follow weak blobs on the steep flanks of the lines, with stochastic errors putting the odd point outside the allowed area.

\subsubsection{Cross-correlation Analysis} \label{322}

\begin{figure}
	\epsfxsize=8cm\epsfbox[40 190 560 690]{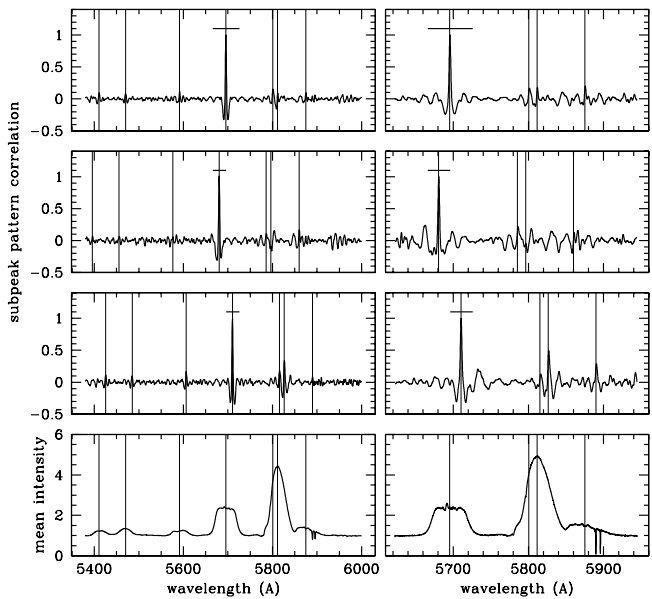}
	\caption{Results from the cross-correlation analysis of the
variability patterns observed in the {CIII} $\lambda5696$ emission
line. The left panel shows the 1999 multi-site campaign (OHP, OMM, DAO, TO, RO) results
while the right panel shows the results obtained for the KPNO run.
Local maxima indicate regions where the variability patterns are
correlated to those detected in {CIII} $\lambda5696$ (note the value of
1.0 reached at 5696 {\AA} indicating maximum correlation of the pattern
with itself). Vertical lines mark positions of the most
prominent emission lines relative to the part of {CIII} $\lambda5696$ used for the correlation. Upper panels show
the correlation functions calculated using the complete pattern in the {CIII} $\lambda5696$ line
($5676$$<\lambda<$$5716$). The maxima are higher when one uses
the patterns from the red side ($5696$ $<\lambda<$ $5716$) of the {CIII} line (bottom correlation panels).
The correlation essentially vanishes when one uses the blue-side patterns, 5676$<\lambda<$5696 (middle correlation panels).
	\label{fig:f11}}
\end{figure}

We now verify if the variable subpeak patterns in the different
emission lines are similar. A visual examination of the spectroscopic
time series (Figs. \ref{fig:f2}, \ref{fig:f4}) does
suggest that the variability pattern in different lines may indeed be similar,
i.e. that specific narrow subpeaks show up in each line, at the same
position of the line profile, and at any given time.
To quantify the similarity in the subpeak patterns, we apply the
cross-correlation technique developed by \citet{lep2000}. The
well-defined variable subpeak pattern in the {CIII} $\lambda5696$
line is used as a reference (the correlation response). Properly normalized,
perfect match of the pattern
at all times would yield a correlation value of $1.0$, while the absence of
any similarity should yield a correlation value close to $0.0$, and
$-1.0$ for an anti-correlation. Results are independent of the
intensity of the pattern; hence if a pattern is identical to the
reference pattern, it will yield a high correlation value, even if its
amplitude is only a fraction of that of the reference pattern. One
caveat is that instrumental noise can significantly degrade a subpeak
pattern, especially if its intensity is low. Since the variable
subpeak patterns in {HeII} $\lambda5411$, {CIV} $\lambda5471$
and {OV} $\lambda5592$ are near the noise limit, one should not
expect them to be highly correlated with {CIII}
$\lambda5696$. However, a small, but clear correlation should be
regarded as significant.

In order to minimize the degradation in the correlation because of
instrumental noise, we apply the cross-correlation technique to the
wavelet-filtered spectroscopic time series (Figs.
\ref{fig:f2}, \ref{fig:f4}). Results of the
cross-correlation analysis are displayed in Figure
\ref{fig:f11}. The cross-correlation is performed separately on
the KPNO time series, compared to the multi-site (OHP, OMM, DAO, TO, RO)
time series. As in \citet{lep2000}, we use three different parts of
the {CIII} $\lambda5696$ line as the reference. First,
we use the complete pattern in the {CIII} $\lambda5696$ line ($5676$$<\lambda<$$5716$)
 as our correlation response (Fig. \ref{fig:f11}, top).
As expected, we find a correlation value of $1.0$ at the position of
the {CIII} $\lambda5696$ line itself. We also find secondary maxima
{\em at the exact positions of all the most prominent emission
lines}. Although some of these maxima are quite
weak (with a correlation value of $\approx0.1$) they appear to be
statistically significant because they occur at the exact wavelength
positions expected for these atomic lines. The low correlation coefficient is a direct consequence
of the weakness (proximity to the noise level) of the variability patterns in all the lines, except in {CIII}
$\lambda5696$. It is also possible that the variability patterns in the
different lines are not exactly the same, as, generally,  the lines
do not form in the same parts of the wind.

We repeat the cross-correlation analysis, this time using only half
of the variability pattern, either on the blue
($5676$$<\lambda<$$5696$) or red ($5696$$<\lambda<$$5716$) side of the
{CIII} $\lambda5696$ emission line (Fig. \ref{fig:f11},
second and third panels from the top). Results show a much stronger correlation when one
uses the red side of the emission line. This perfectly mimics the behavior of the subpeaks in
the presumably single star WR135 \citep{lep2000}. Most WR
emission lines are flanked on their blue side by a weak {\em but variable} P-Cygni absorption 
trough, with a notable exception of the
{CIII} $\lambda5696$ emission line. Generally, the P-Cygni absorption trough
is the most variable part of the profile. This may
introduce an extra component of variability on the blue edge of the
line. However, because {CIII} $\lambda5696$ is unaffected by
P-Cygni variations, the intrinsic variability of its blue edge is
significantly different from that in the other lines. This explains
the lower correlation between the blue edge of {CIII}
$\lambda5696$ and the blue edge of the other lines.

\subsection{Radial velocity variations: Orbit of the two components}

\begin{figure}
      \epsfxsize=7cm\epsfbox[50 190 480 610]{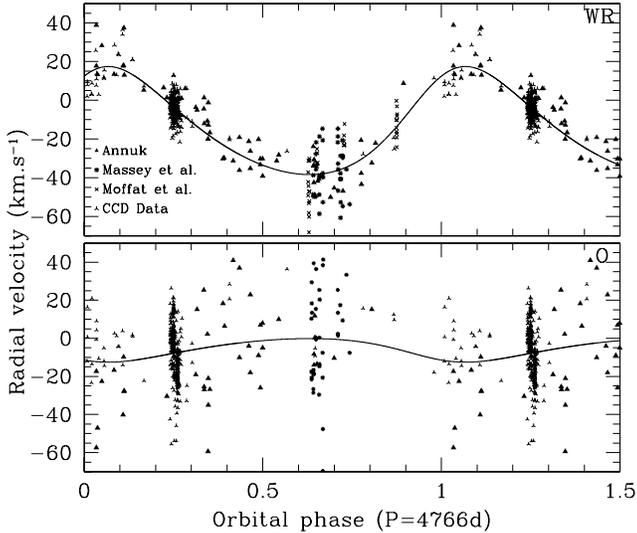}
       \caption{{Radial velocities of the WR component phased with
	the 4766d orbit are shown in the upper panel: data from \citet{ann1991,ann1995}, \citet{mas1981} and
	\citet{mof1986} refer to photographic plates; the rest are measured from the current
	CCD spectra from DAO, OMM, OHP, RO, TO and KPNO. A typical standard deviation for the measurements
	is $\sim$ $15$ $km\,s^{-1}$ for the former and $10$ $km\,s^{-1}$ for the latter. The bottom panel shows radial
	velocities for the O component. Full lines show the derived orbit.}
        \label{fig:f12}}
\end{figure}

Radial velocities (RV) were measured by fitting gaussian profiles to
the {HeI} $\lambda5876$ absorption line corresponding to the O star and cross-correlating the template
(an average constructed from the data sets coming from different observatories) with the individual spectra in the region
covering the strong {CIV} $\lambda\lambda5802/12$ emission blend for the WR component.
Typical uncertainty of the measurements run at $\sim$ 5 - 10 $km\,s^{-1}$ for an individual spectrum of the WR star and 
$\sim$ 10 - 20 $km\,s^{-1}$ for the O component.
The results were corrected for systematic differences between observation sets.
This adjustment was done taking the RO and OMM observations together,
as in the overlapping epochs they show negligible relative shifts for the WR star measurements:
the average velocity measured on the overlapping sub-sets (summer 1999) was
$<{V_{RO}}>$ = $3.0 \pm 6.7$ $km\,s^{-1}$,  $<{V_{OMM}}>$ = $3.6 \pm 5.6$ $km\,s^{-1}$.
Then, we shifted the Kitt peak RV data into the RO and OMM system and applied the same procedure to the less
numerous DAO and OHP sets (shifted relative to the RO+OMM system by $7-9$ $km\,s^{-1}$).
The Tartu set was not modified, as many of its data were taken later on, without any substantial overlap (see Table~\ref{tbl3}, published
separately in the electronic version).

\begin{table*}
\centering
\begin{minipage}{110mm}
\caption{Radial velocities of the WR and O components of WR137.\label{tbl3}}
\begin{tabular}{cccccc}
\hline
{HJD-    } & {RV} & {HJD-    }     & {RV} & {HJD-    }     & {RV} \\
{\hskip -4.5mm  2440000} & { $km\,s^{-1}$  } &  {\hskip -4.5mm  2440000} &
{ $km\,s^{-1}$  }   & {\hskip -4.5mm  2440000}&  { $km\,s^{-1}$  } \\    
\hline
{\large {\bf WR}}& {\large {\bf Comp.}}& &   & $\sigma$$\sim$ 5-10 $km\,s^{-1}$& \\ 
\hline
 {\bf KPNO}   &  n=58    &    {\bf OHP}   &     n=50       &   {\bf RO}	     & n=67		\\
\hline
11432.63091  & 0.2    &   10836.23647  &    21        &   11337.86891	     & -9.2		  \\
11432.66730  &  1.1   &   10837.24652  &    21        &   11350.80538	     &   -5.1		    \\
11432.74425  &  3.5   &   10461.23148  &    15.54     &   11355.78767	     &   -13.4  	    \\
11432.76590  &  4.7   &   10454.24317  &    11.9      &   11364.71949	     &   0.6		    \\
11432.78742  &  5.6   &   10456.22850  &    11.15     &   11364.84482	     &   -11.1  	    \\
\hline
\end{tabular}

\medskip	
{ Table~\ref{tbl3} is published in its entirety in the electronic edition of this article.
A portion is shown here for guidance regarding its form and content. {\rm n} corresponds to the total number of
spectra for one observatory.}
\end{minipage}
\end{table*}

With the assumption that our data set alone may not be adequate in
revealing the presumably long orbital period of WR137, we have incorporated earlier published RV measurements
\citep{mas1981,ann1991,ann1995,mof1986}. We applied the orbit-fitting algorithm 
from \citet{ber1969}, assigning  individual $\gamma$-velocities for the following sets:
all the new CCD data for the WR component, as discussed above (made homogeneous by additional adjustments),
then the WR sets of \citet{mas1981} and  \citet{mof1986} (photographic data, treated as 2 individual sub-sets), then 2
sets for the O star, one from our campaign, and the second described in \citet{mas1981}.
This allowed us to compensate for the systematic differences in the RVs coming from different systems and epochs and
evaluate the overall accuracy of the data:
for the CCD data, $\sigma_{o-c}$ is $\sim$ $6$ $km\,s^{-1}$ for the orbital fit of the WR component
while it is $\sim$ $15$ $km\,s^{-1}$ for the O component, in line with the accuracies attained in measurements of
individual spectra. The photographic plates yielded
$\sigma_{o-c}$ $\sim$ $11$ $km\,s^{-1}$ and $\sim$ $26$ $km\,s^{-1}$ for the WR and O components, respectively.
Despite the different accuracy of the photographic and CCD data, we have kept the same weights while deriving orbital
elements: less accurate photographic data have, nevertheless, an advantage of being taken at different epochs, while
the CCD set mainly concentrates around 1998-2000, which is not very helpful in the situation when the orbital period exceeds 10 yrs. 

The orbital elements obtained from the least-square fit of {\it 427} WR RVs and {\it 373} O RVs
are listed in Table~\ref{tbl4}. Figure \ref{fig:f12} shows that the computed orbit based on these parameters
fits quite well all RV data obtained so far for the WR component. On the other hand, the fit for the O component is less
precise, due to relatively large errors of the measurements.

\begin{table}
\caption{Orbital elements of HD192641 \label{tbl4}}
\begin{tabular}{@{}lccc}
\hline
Parameter & WR   & O   \\
\hline
  $P$ $(d)$ & 	\multicolumn{2}{c}{  $4766 \pm 66 $ 	} \\
$T_{0} (JD)$ &	\multicolumn{2}{c}{ $2450198 \pm 186 $ } \\
$\omega$ ($^{\circ})$& \multicolumn{2}{c}{ $326 \pm 15$ }\\
$e$   & \multicolumn{2}{c} {$0.178 \pm 0.042$ }	 \\
 $\gamma$ ($km\,s^{-1}$)  	& $-14.6 \pm 0.9$ & $-5.4 \pm  0.7$	 \\
 $K$ ($km\,s^{-1}$) &  $27.9 \pm 1.2 $&  $6.1 \pm 1.3$	 \\
$\sigma_{CCD}(o-c)$ ($km\,s^{-1}$) & 6 & 15 \\
$\sigma_{plates}(o-c)$ ($km\,s^{-1}$) & 11 & 26 \\
$a sini$ ($AU$)  & $12 \pm 0.5$ & $2.7 \pm 0.6$ 	 \\
$M sin^{3}i$ ($M_{\odot}$) & $3.4 \pm 1.0$  & $15.3 \pm 2.1$ \\
 \hline
\end{tabular}

\medskip
{These parameters were derived with the combined radial velocities of
CIV $\lambda$5806.0 and absorption at HeI $\lambda$5875.6 using
the program developed by \citet{ber1969}.}
\end{table}

\section{Discussion and conclusions}

\subsection{Spectral line variability}
The presence of propagating stochastic structures indicates a clumpy wind as 
in all WR stars appropriately studied so far. Moreover, this study enabled us to constrain 
the parameters of the assumed $\beta$-law with good accuracy, due to 
the large quantity of data. We found $\beta R_{\star} \simeq 25 \pm 5 R_{\odot}$
(according to Figure \ref{fig:f9}) which agrees marginally well with previous results, 
$\sim$ 35-90 $R_{\odot}$ \citep{lep1999}, found using a wavelet analysis for WR137.
However, one should be cautioned against any oversimplified interpretation of the $\beta$-law. \citet{ham2001}
have discussed the apparent contradiction between the fast velocity law ($\beta$ $\sim$ $1$) and
the slow drift of the DACs \citep[Discrete Absorption Components: ][]{cra1996} in the winds of O stars.
The explanation resides either in a difference between the velocity of patterns (DACs) and structures (modulations) in
the wind or in different velocity laws co-existing at the same time in different spatial zones
of the wind.
Indeed, one may consider alternatives  to the original $\beta$-law. E.g., \citet{hil1999} introduced two power indices,
thus dividing the wind into two distinct parts with different expansion laws.
Although such a law might fit real data better, it is really no more physically justified than the simple $\beta$-law.

In addition, wavelet analysis and cross-correlation yield other interesting results. First,  although weak, there is
a correlation in variability patterns between all the most prominent lines in the $\lambda \lambda 5300-6000$ {\AA} range.
Moreover, this correlation is stronger for the red-side than the blue-side variability patterns of {CIII} $\lambda5696$: 
in the region of interest  most emission lines -except the strong
{CIII} $\lambda5696$ line-  show variable P-Cygni absorption troughs, which efficiently decorrelates the observed patterns.
This behaviour was also detected in spectra of the single WC8 star WR135 \citep{lep2000}.

\subsection{Short period} \label{disc}

 The 0.83d$\pm$0.04d period could be related to large-scale, coherent  structuring
 of the WR wind, taking the form of CIRs \citep[Corotating Interaction Regions: ][]{cra1996}.
 CIRs are thought  to be responsible for the periodic, P=3.766d, patterns re-appearing in the wind of WR6
 \citep[their Fig. 5]{stl1995}, with a coherency timescale of at least $\sim$ 4 rotation cycles. 
 
If this 0.83d period was related directly to a phenomenon occuring on the surface of the star, it could imply a rotational velocity that is closer 
to the break-up velocity in WR137 than in WR6 or WR134. Taking parameters from Table~\ref{tbl2}, we obtain a rotational velocity of 275 $km\,s^{-1}$ and 
a break-up velocity of 475 $km\,s^{-1}$, which give $V_{rot}/V_{break-up}$ $\simeq$ $0.58$. The results are not inconsistent with a rotational period, 
although one might also want to look for other explanations.
This period could be only a fraction of the real rotation period, as is often supposed for CIRs \citep{stl1995} with multiple arms. Moreover, relatively rapid 
rotation of WR137 could be related to a flattened wind, as implied by the depolarized emission lines \citep{har2000}.

The O star companion with its period of 4766 days cannot possibly account for such short-term variability, as the 0.83d
period is seen exclusively in the absorption troughs of the lines originating in the WR wind.
The existence of a much closer, low-mass compact companion could be another explanation.
However,  WR137 cannot be considered as a bright X-ray source:
$L_{X}(0.2-2.4keV)$$=$$0.25 \pm 0.14$ $10^{32}$ $erg$ $s ^{-1}$ \citep{pol1995}.
On the other hand, the absence of a high X-ray output  (compared to Cyg X-3 at $\sim 10^{38} erg s^{-1}$) does not necessarily
mean that there is no compact companion \citep{erg1998,mar1996} in so far as the {\it propeller}
effect created by the rotating magnetic field may inhibit production of X-rays. A period of 0.83 days would imply an orbital
separation $r_{O}$ $\simeq$ $7$ $R_{\odot}$ ($6.7\pm 0.8$ $R_{\odot}$) between the WR star and
a hypothetical companion, according to the mass for the WR star (see Table~\ref{tbl2}) and considering a neutron star of $1.4$ $M_{\odot}$.
Since $R_{\star}$ $\simeq$ $4.5$ $R_{\odot}$, this hypothesis remains plausible until further evidence
can be found to the contrary.\\
\newfont{\sss}{cmmi12 scaled 1200}
Another possible source of the 0.83d variability could be related to pulsations of the WR star. Periods of radial pulsations in 
WR stars are predicted to range
from several minutes to an hour \citep{bra1996} and thus can be ruled out. The same holds true for strange-mode pulsations \citep{saio1998}, 
for which expected periods are more likely to be in the range of  several minutes \citep{scha1992,gla1993}.
 The claim of P $\sim 627s$ for the southern Wolf-Rayet star WR40 \citep{ble1992} was contradicted by the results of
\citet{mar1994} and \citet{sch1994}, who claimed this period could be either spurious (caused by atmospheric transparency variations) 
or a pulsation appearing and vanishing on timescales of hours to a day (although no build-up or fading was noticed in the intensity of the variation).
 On the other hand, {\it non-radial} pulsations with a period $\sim$ $20$ hours ($0.83$d) are a possibility. They could be driven by the
 Eddington $\epsilon$-mechanism \citep{mae2002,edd1918}. The timescales of these modes would be of the order of
 hours to less than a day \citep{noe1986,dzi1994} so that they could be held accountable for the observed period in WR137. These are related 
 to g+ modes of non radial pulsation, but are not believed to last more than a few $10^{3}$ years. Further studies of WR137 with good temporal and spectral resolution (comparable to the current KPNO spectra) and high signal to noise
ratio (S/N $\simeq$ 100-200) could help to confirm and determine the origin of this small amplitude periodic variability. 
Note that the non-radial g-mode oscillations should have a variable period \citep{bal1989}, while the rotationally-modulated variability 
should provide a stable periodic modulation, thus providing means to distinguish between the two.

\subsection{Orbital period and masses of the components}
We have computed a reliable orbit for the WR star and provided the best possible estimate for the O-component orbit.
Our derived period of $4766 \pm 66$ days is in perfect agreement with that
found independently from the near-IR photometric data taken between 1985 and 1999 \citep{wil2001}, which yielded $4765 \pm 50$ days.
The small eccentricity ($e=0.178$) seems to be compatible with the episodic dust formation depicted in \citet{mar1999}, who
assumed $e \le 0.3$, based on the rather
slow variability of the IR flux \citep{wil2001} and behavior of  the  ejected dust cloud. Assuming that we see the system fairly near to edge-on,
this would imply that the dust clumps observed
in 1997 September - 1998 May with HST/NICMOS2 are indeed formed around the time when the O-companion was relatively close to the
WR star. 
Indeed, the periastron passage seems to occur around 1996 (between February and October), according
to Table~\ref{tbl4}.
Assuming a distance to the system of $1.82$ kpc (Table~\ref{tbl2}) and a spatial ejection velocity of $\sim$ 2000 $km\,s^{-1}$ \citep{mar1999},
the displacement of the persistent dust cloud by $0.25$" in May 1998 implies that the cloud was formed during periastron passage in
1996, long before the 1997-1998 images
were taken. The diminishing orbital separation around periastron should intensify the wind-wind collision and, finally,
increase the dust output (e.g. like another long-period dust-producing binary WR 140: \citet{mar2003}).
One more encouraging detail is the orientation of the WR orbit. Indeed, the behavior of the dust cloud implies that it was ejected
from the system practically in the plane of the sky, at $\alpha_{dust}\sim 27^o$ relative to the plane \citep{mar1999}. This fits well the longtitude of
periastron passage of the O star: $326^o-180^o=146^o$ (Table~\ref{tbl4}) which implies $\alpha_{orbit}=180^o-146=34^o$, thus $\alpha_{orbit} \sim \alpha_{dust}$.

The \citet{huc2001} catalogue provides an absolute magnitude of Mv=-5.7 for the WC7pd+O9 system, and Mv=-4.52 for the WR-component, 
implying Mv=-5.25 for
the O-component. Thus, according to \citet{vac1996}, the O9 component should have a spectral class between III and V and a
mass of $M_{O}$=20$\pm$2 $M_{\odot}$ \citep{gie2003}. This,
in turn, implies $M_{WR}$=4.4$\pm$1.5 $M_{\odot}$ and an inclination angle of i=$67^{\circ}$.
On the other hand, \citet{nug2000} give a mean of $M_{WR}$=13.4 $\pm$ 1.3 $M_{\odot}$ for WC7 stars which, if assumed to be applicable
to the WC7 star in WR137, leads to  $M_{O}$=60.9 $\pm$ 15.5 $M_{\odot}$
and i=$39^{\circ}$. However, the shape of the ejected dust
cloud strongly favors the larger value of inclination. Moreover,  the mutual orientation of the orbital plane and
the plane containing the flattened wind \citep{har2000} also points to a rather high value of $i$.

\section*{Acknowledgments}
{\it We wish to thank the Observatoire de Strasbourg for its support to LL's PhD thesis. AFJM thanks NSERC (Canada) and FQRNT (Qu\'ebec) for
financial support.}

\clearpage

\label{lastpage}
\end{document}